\documentclass[12pt]{iopart}

\usepackage[dvips]{epsfig}

\begin{document}

\title[Dobi\'nski-type relations: Some properties and physical applications]
{Dobi\'nski-type relations: Some properties and physical applications}
\author{P Blasiak$^{a}$, A Horzela$^{a}$, K A Penson$^{b}$
and A I Solomon$^{b,c}$\vspace{2mm}}

\address
{$^a$ H. Niewodnicza\'nski Institute of
Nuclear Physics, Polish Academy of Sciences\\
ul. Eliasza-Radzikowskiego 152,  PL 31342 Krak\'ow, Poland\vspace{2mm}}

\address
{$^b$ Laboratoire de Physique Th\'eorique de la Mati\`{e}re Condens\'{e}e\\
Universit\'e Pierre et Marie Curie, CNRS UMR 7600\\
Tour 24 - 2i\`{e}me \'et., 4 pl. Jussieu, F 75252 Paris Cedex 05, France\vspace{2mm}}

\address
{$^c$ The Open University, Physics and Astronomy Department\\
Milton Keynes MK7 6AA, United Kingdom\vspace{2mm}}

\eads{\linebreak \mailto{pawel.blasiak@ifj.edu.pl}, 
\mailto{andrzej.horzela@ifj.edu.pl}, 
\mailto{penson@lptl.jussieu.fr}, 
\mailto{a.i.solomon@open.ac.uk}}


\begin{abstract}
\\
We introduce a generalization of the Dobi\'nski relation through which 
we define a family of Bell-type numbers and polynomials. 
For all these sequences we find the weight function of the moment problem 
and give their generating functions.
We provide a physical motivation of this extension in the context of
the boson normal ordering problem and its relation to an extension of
the Kerr Hamiltonian.
\end{abstract}

\maketitle

In this note we consider the Dobi\'nski relation and its generalization. 
This topic naturally belongs to the field of combinatorial analysis. 
The Dobi\'nski relation \cite{Dobinski} was first derived in connection with 
\emph{Bell numbers} $B(n)=1,1,2,5,52,203,877,\ldots$, $n=0,1,2,...$, which describe partitions of a set \cite{Bell},\cite{Rota}.
That remarkable formula represents the integer sequence $B(n)$ as an infinite sum of ratios 
\begin{eqnarray}\label{BellDob}
B(n)=e^{-1}\sum_{k=0}^\infty \frac{k^n}{k!}.
\end{eqnarray}
Closely related to the Bell numbers are \emph{Stirling numbers} of the second kind $S(n,k)$, $k=1...n$, and the \emph{Bell polynomials} defined as
\begin{eqnarray}
B(n,x)=\sum_{k=1}^n S(n,k)\,x^k,
\end{eqnarray}
related to $B(n)$ by $B(n)=B(n,1)=\sum_{k=1}^n S(n,k)$.
For the Bell polynomials the Dobi\'nski relation Eq.(\ref{BellDob}) generalizes to
\begin{eqnarray}\label{BellDobX}
B(n,x)=e^{-x}\sum_{k=0}^\infty \frac{k^n}{k!}\,x^k.
\end{eqnarray}
These formulas may be derived using either combinatorial or purely analytical methods  
starting from the original interpretation of Bell and Stirling numbers given in enumerative combinatorics \cite{Comtet},\cite{Wilf}. Accordingly, 
Stirling numbers $S(n,k)$ count the number of possible partitions of the $n$ element set into $k$ subsets (none of them empty) and
Bell numbers $B(n)$ count all such partitions.
We note that other pictorial representations can also be given, {\it e.g.}
in terms of graphs \cite{Mendez} or rook numbers
\cite{Navon},\cite{QTS3},\cite{Varvak}.

One may conversely take Eq.(\ref{BellDob}) (or Eq.(\ref{BellDobX})) as the definition of the Bell numbers (or polynomials).
This observation suggests the generalization of these sequences through the 
Dobi\'nski relation. In this note we introduce an extension of Eq.(\ref{BellDob})
and define the family of \emph{Bell-type numbers} as
\begin{eqnarray}\label{Dob}
\mathcal{B}(n)=\sum_{k=0}^\infty \frac{\left[P(k)\right]^n}{D(k)},
\end{eqnarray}
where $P(k)$ and $D(k)$ are any functions of $k=0,1,2,...$ such that $D(k)\neq 0$ and the above sum converges.
Note that conventional Bell numbers are obtained for $P(k)=k$ and $D(k)=e\,k!$ .
\\
This generalization was also pointed out in \cite{Myczkowce} and \cite{BPSLogNormal} in connection with the log-normal distribution.
Here we focus on the general properties of our proposed definition and show that the very specific form of Eq.(\ref{Dob}) results in a straightforward solution of the moment problem and calculation of the generating functions. We also comment on the connection to physics and interpret the  sequences so defined in the context of the problem of the normal ordering of boson operators.

\begin{figure}
\hspace{1cm}
\resizebox{13cm}{!}{\includegraphics{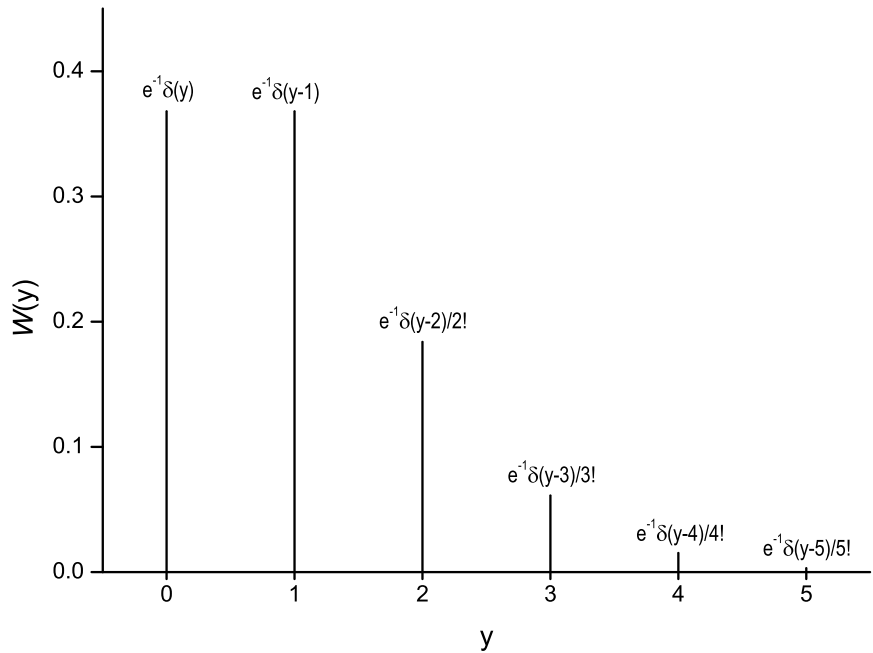}}
\caption{\label{FigBell}The portion for $0 \leq y \leq 5$ of the weight function $W(y)=e^{-1}\sum_{k=0}^\infty \frac{\delta(y-k)}{k!}$ solving the Stieltjes moment problem for the Bell numbers $B(n)$. Height of the vertical lines is proportional to the weight of the Dirac $\delta$ functions.}
\end{figure}

Suppose that  we want to solve the moment problem \cite{Akhiezer} for the sequence $\mathcal{B}(n)$, \emph{i.e.} we seek a positive weight function $\mathcal W(y)$ such that $\mathcal{B}(n)$ is its $n$-th moment
\begin{eqnarray}\label{M}
\mathcal{B}(n)=\int dy\ y^n\,\mathcal W(y).
\end{eqnarray}
At this point  we do not specify the domain of $\mathcal{W}(y)$ or the limits of the integral.
A closer look at Eq.(\ref{Dob}) yields the following candidate for the weight function 
\begin{eqnarray}\label{W}
\mathcal W(y)=\sum_{k=0}^\infty \frac{\delta(y-P(k))}{D(k)}.
\end{eqnarray}
This is an infinite ensemble of weighted Dirac $\delta$ functions located at a specific set of points $\{P(k),\, k=0,1,2,...\}$ and is called a \emph{Dirac comb}. If all the weights $1/D(k)$ are positive ($D(k)>0$) and normalized to one ($\sum_{k=0}^\infty 1/D(k)=1$) then Eq.(\ref{W}) is a positive and normalized distribution which is a solution of the moment problem of Eq.(\ref{M}). Whether it corresponds to the Hamburger, Stieltjes or Hausdorff moment problem depends on the range of the set $\{P(k),\, k=0,1,2,...\}$.
For example, for the sequence of Bell numbers $B(n)$ the weight function $ \mathcal W(y)=e^{-1}\sum_{k=0}^\infty \frac{\delta(y-k)}{k!}$ is a positive and normalized distribution solving the Stieltjes moment problem, see Fig.\ref{FigBell}. A solution of the Hamburger moment problem is generated by the set of \emph{restricted} Bell numbers $B_{\overline{1}}(n)=1,0,1,1,4,11,41,162,\ldots$ for $n=0,1,\ldots$ counting partitions without singletons \cite{Comtet}. They satisfy $B_{\overline{1}}(n)=e^{-1}\sum_{k=0}^\infty\frac{(k-1)^n}{k!}$ with $P(k)=k-1$ and $D(k)=e\,k!$; the measure is ${\mathcal W}_{\overline{1}}(y)=e^{-1}\sum_{k=0}^\infty \frac{\delta(y-k+1)}{k!}$. On the other hand, the well-known Catalan numbers 
$C(n) =\frac{1}{n+1}\frac{}{}\left(\!\!\begin{array}{c}2n\vspace{-2mm}\\n \end{array}\!\!\right)$ are solutions of the Hausdorff moment problem \cite{psix}.

The specific form of Eq.(\ref{Dob}) simplifies the calculation of the generating functions. Taking the {\em exponential} generating function, substituting Eq.(\ref{Dob}) and changing the summation order one obtains
\begin{eqnarray}\label{egf}
\mathcal{G}(\lambda)=\sum_{n=0}^\infty \mathcal{B}(n)\frac{\lambda^n}{n!}=\sum_{k=0}^\infty \frac{e^{\lambda P(k)}}{D(k)}.
\end{eqnarray}
Evaluation of this series depends on the particular choice of the functions $P(k)$ and $D(k)$ and in general the series may be divergent. For the Bell and restricted Bell numbers Eq.(\ref{egf}) it can be evaluated easily: $G(\lambda)=\sum_{n=0}^\infty B(n)\frac{\lambda^n}{n!}=e^{e^\lambda -1}$ and $G_{\overline{1}}(\lambda)=\sum_{n=0}^\infty B_{\overline{1}}(n)\frac{\lambda^n}{n!}=e^{e^\lambda -1-\lambda}$.
\\
Similarly, for the {\em ordinary} generating function one gets
\begin{eqnarray}\label{ogf}
\mathcal{G}_o(\lambda)=\sum_{n=0}^\infty \mathcal{B}(n)\,\lambda^n=\sum_{k=0}^\infty \frac{1}{D(k)\cdot(1-P(k)\,\lambda)}.
\end{eqnarray}
The same procedure can also be performed  for other cases, \emph{e.g.} for hypergeometric generating functions \cite{BPSPLA}. The choice of the denominator in the generating function may depend on $P(k)$, $D(k)$ and the purpose we need it for (\emph{e.g.} when we need a well convergent generating function for analytical calculations).

In the same manner one could generalize Eq.(\ref{BellDobX}) and define 
\begin{eqnarray}\label{DobX}
\mathcal{B}(n,x)=\sum_{k=0}^\infty \frac{\left[P(k,x)\right]^n}{D(k,x)}.
\end{eqnarray}
The additional variable in the functions $P(n,x)$ and $D(n,x)$ does not pose any complication either in the proposed approach to the moment problem or in the evaluation of the generating functions.
However, we must observe that in general $\mathcal{B}(n,x)$ has an infinite expansion in $x$ and only for particular choices of the functions $P(k,x)$ and $D(k,x)$ does it yield {\em polynomials}. It is certainly the case for $D(k,x)= k!\,e^x\,x^{-k}$, and $P(k,x)=P(k)$ a polynomial in $k$. Therefore we define the \emph{Bell-type polynomials} as
\begin{eqnarray}\label{DobXX}
\mathcal{B}(n,x)=e^{-x}\sum_{k=0}^\infty \frac{\left[P(k)\right]^n}{k!}x^k.
\end{eqnarray}
As a result the weight function of Eq.(\ref{W}) takes the form 
\begin{eqnarray}\label{WX}
\mathcal W(x,y)=e^{-x}\sum_{k=0}^\infty \frac{\delta(y-P(k))}{k!}x^k,
\end{eqnarray}
and the exponential generating function of Eq.(\ref{egf}) is 
\begin{eqnarray}\label{egfX}
\mathcal{G}(\lambda,x)=e^{-x}\sum_{k=0}^\infty \frac{e^{\lambda P(k)}}{k!}x^k.
\end{eqnarray}
The case of conventional Bell polynomials is obtained for $P(k,x)=k$. Consequently one gets the positive and normalized weight function $\mathcal W(y,x)=e^{-x}\sum_{k=0}^\infty \frac{\delta(y-k)}{k!}\,x^k$ of the Stjelties moment problem, $B(n,x)=\int_0^\infty dy\ y^n\, \mathcal W(y,x)$, and the exponential generating function $G(\lambda,x)=\sum_{n=0}^\infty B(n,x)\frac{\lambda^n}{n!}=e^{x(e^\lambda -1)}$ (see \cite{HierDob}). Analogous considerations can be applied to the polynomials generated by $B_{\overline{1}}(n)$ leading to $B_{\overline{1}}(n,x)=e^{-x}\sum_{k=0}^\infty \frac{(k-1)^n}{k!}x^k$, and $G_{\overline{1}}(\lambda,x)=\sum_{n=0}^\infty B_{\overline{1}}(n,x)\frac{\lambda^n}{n!}=e^{x(e^\lambda -1-\lambda)}$.

Introduction of the generalized Bell-type numbers and polynomials through Eqs.(\ref{Dob}) and (\ref{DobXX}) is not merely a mathematical
definition but it has a firm grounding in physics. We will show that it is related to the solution of the normal ordering problem for a general function of the number operator.
\\
Consider the boson creation operator $a^\dag$ and annihilation operator $a$  satisfying the commutator $[a,a^\dag]=1$. Suppose we are given a  function of these operators. 
Its {\em normally ordered} form is obtained by moving all the creation operators to the left of the annihilation operators using the  commutation relation. The normal ordering procedure is of fundamental importance in quantum mechanical calculations in the coherent state representation, the latter defined
by the coherent states $|z\rangle=e^{-|z|^2/2}\sum_{n=0}^\infty \frac{z^n}{\sqrt{n!}}|n\rangle$, where $a^\dag a |n\rangle =n|n\rangle $, $\langle n|n'\rangle =\delta_{n,n'}$ and $a |z\rangle =z|z\rangle $ \cite{Klauder}.
For example, if we take the $n$-th power of the number operator $a^\dag a$ the normal ordering procedure gives \cite{Katriel}
\begin{eqnarray}\label{aa}
(a^\dag a)^n=\sum_{k=1}^n S(n,k)\,(a^\dag)^k a^k.
\end{eqnarray}
It involves Stirling numbers of the second kind $S(n,k)$ and the coherent state matrix element yields the Bell polynomial
\begin{eqnarray}
\langle z|(a^\dag a)^n|z\rangle=B(n,|z|^2).
\end{eqnarray}
Now we consider a general polynomial of the number operator denoted by
\begin{eqnarray}\label{H}
    \mathcal{H}_{\alpha}(a^\dag a) =\sum_{k=N_0}^N \alpha_k (a^\dag a)^k
\end{eqnarray}
with some constants $\alpha_k$; $N_0$ and $N$ are the smallest and
largest indexes of non-vanishing $\alpha_k$, respectively. Physically,
$\mathcal{H}_{\alpha}$ may be thought of  as a generalisation of the
Kerr Hamiltonian of quantum optics \cite{Kerr}. 
The $n$-th power of $\mathcal{H}_{\alpha}$ defines the \emph{Stirling-type numbers} as
\begin{eqnarray}\label{Hn}
\left[\mathcal{H}_{\alpha}(a^\dag a)\right]^n=\sum_{k=N_0}^{nN} \mathcal{S}_{\alpha}(n,k)\ (a^\dag)^ka^k,
\end{eqnarray}
and associated \emph{Bell-type polynomials} (of order $nN$) are
\begin{eqnarray}\label{B}
\mathcal{B}_{\alpha}(n,x)=\sum_{k=N_0}^{nN}\mathcal{S}_{\alpha}(n,k)\,x^k.
\end{eqnarray}
We will show now that such polynomials defined in the normal ordering problem correspond to the Bell-type polynomials introduced in Eq.(\ref{DobXX}).
\\
To this end observe that $[a,a^\dag]=[D,X]=1$ where $D$ and $X$ 
are the derivative and multiplication operators. 
We first rewrite Eq.(\ref{Hn}) in terms of $D$ and $X$:
\begin{eqnarray}\label{HDX}
\left[\mathcal{H}_{\alpha}(XD)\right]^n=\sum_{k=N_0}^{nN} \mathcal{S}_{\alpha}(n,k)\ X^kD^k.
\end{eqnarray}
By acting with the r.h.s. of Eq.(\ref{HDX}) on $e^x$ one obtains $e^x\mathcal{B}_{\alpha}(n,x)$.
Action of the l.h.s. on $e^x$ is more involved. First we apply it to the monomial 
$x^m$ which yields 
$\left[\mathcal{H}_{\alpha} (XD)\right]^nx^m=\left(\sum_{k=N_0}^N\alpha_k\ m^k\right)^nx^m$ from which
$\left[\mathcal{H}_{\alpha} (XD)\right]^ne^x=\sum_{m=0}^\infty \left(\sum_{k=N_0}^N\alpha_k\ m^k\right)^n\frac{x^m}{m!}$ follows.
Combining  these two observations we deduce that
\begin{eqnarray}\label{dobinski}
\mathcal{B}_{\alpha}(n,x)=\sum_{k=N_0}^{nN}\mathcal{S}_{\alpha}(n,k)\,x^k=e^{-x}\sum_{k=0}^\infty \frac{\left[\mathcal{H}_\alpha(k)\right]^n}{k!}\,x^k,
\end{eqnarray}
which has the same form as Eq.(\ref{DobXX}) for $P(k)=\mathcal{H}_\alpha(k)$. This gives an  interpretation of the Bell-type polynomials and numbers  in the context of the normal ordering problem. The assumption that $\mathcal{H}_\alpha(x)$ is a polynomial guarantees that $\mathcal{B}_\alpha(n,x)$ is also a polynomial in $x$. Although this additional assumption may be irrelevant in general, as we have mentioned above it leads to infinite sequences of Stirling-type numbers.
\\
Having solved the normal ordering problem and identified the solution in the framework of the Dobi\'nski-type relations we may now exploit its advantages. The solution of the moment problem of Eqs.(\ref{M}) and (\ref{W}) is immediate (depending on the coefficients $\alpha_k$ in $\mathcal{H}_\alpha(x))$.
Also the generating function setting is easily applicable. We just mention that the normally ordered exponential of a function of the number operator is the exponential generating function of the associated Bell-type polynomials. In the coherent state representation it may be written as
\begin{eqnarray}
\langle z|e^{\lambda\mathcal{H}_\alpha(a^\dag a)}|z\rangle=\mathcal{G}_\alpha(\lambda,|z|^2)=e^{-|z|^2}\sum_{k=0}^\infty \frac{e^{\lambda \mathcal{H}_\alpha(k)}}{k!}\,|z|^{2k}.
\end{eqnarray}
This and other consequences of the normal ordering of polynomial-type Hamiltonians in the number operator will be discussed elsewhere.

In conclusion we want to emphasize the advantages of introduction of the Bell-type numbers and polynomials through generalization of the Dobi\'nski relation of Eqs.(\ref{Dob}) and (\ref{DobXX}). It enables a straightforward solution of the moment problem in the form of a Dirac comb, see Eqs.(\ref{M}), (\ref{W}) and (\ref{WX}). Moreover, calculation of the generating functions simplify considerably in that framework (see Eqs.(\ref{egf}), (\ref{ogf}) and (\ref{egfX})). We have also pointed out that this generalization has immediate application to the boson normal ordering problem. We have interpreted a wide class of Stirling-type numbers as the expansion coefficients of  normally ordered functions of the number operator.

Further modifications of the structure of the infinite series of Eqs.(\ref{Dob}) and (\ref{DobXX}) may, in general, lead to  moment problems with continuous weight functions and will be developed elsewhere.

\ack

We thank M. Ku\'s and M. Yor for important discussions.

\end{document}